\documentclass{ws-ijmpd}
\usepackage[super,compress]{cite}
\usepackage{float}
\begin{document}

\markboth{Carlos A.R. Herdeiro, Eugen Radu \and Helgi F. R\'{u}narsson}
{Spinning boson stars and Kerr black holes with scalar hair: the effect of self-interactions}

\catchline{}{}{}{}{}
%
%%%%%%%%%%%%%%%%%%%%%%%%%%%%%%%%%%%%%%%%%%%%%%%%%%%%%%%%%%%%%%%%%%%%

\title{Spinning boson stars and Kerr black holes with scalar hair:\\ the effect of self-interactions}

\author{Carlos A.R. Herdeiro, Eugen Radu \and Helgi F. R\'{u}narsson}

\address{Departamento de Física da Universidade de Aveiro and Center for Research and Development in Mathematics and Applications (CIDMA)\\
Campus de Santiago, 3810-183, Aveiro, Portugal\\
herdeiro@ua.pt, eugen.radu@ua.pt, helgi.runarsson@ua.pt}

\maketitle

\begin{history}
\received{Day Month Year}
\revised{Day Month Year}
\end{history}

\begin{abstract}
  Self-interacting boson stars have been shown to alleviate the astrophysically low maximal mass of their non-self-interacting counterparts. We report some physical features of spinning self-interacting boson stars, namely 
  their compactness, the occurence of ergo-regions and the scalar field profiles, for a sample of values of the coupling parameter. The results agree with the general picture that these boson stars are comparatively less compact than the non-self-interacting ones.  We also briefly discuss the effect of scalar self-interactions on the properties of Kerr black holes with scalar hair.
\end{abstract}

\keywords{Black holes; Scalar field; Boson stars; Self-interactions}

\ccode{PACS numbers: 04.20.Jb 04.40.-b }

%\tableofcontents

\section{Introduction}	

Boson stars (BSs) are self-gravitating solitons. In their simplest guise, they arise as solutions of General Relativity minimally coupled to a massive, free, complex scalar field. \cite{Kaup:1968zz,Ruffini:1969qy} This simple model presents only two dimensionful parameters (or scales): Newton's constant $G$ (which can be rephrased as the Planck mass $M_{\rm Pl}$) and the scalar field mass $\mu$.  As for fermionic stars, the scalar field effective pressure cannot withstand an infinite amount of mass without undergoing gravitational collapse into a black hole (BH). Intuitively, the smaller the scalar field mass, the larger the maximal mass of the star can be, since the scalar particle's Compton wavelength is larger and hence these particles are less confined. 
Indeed, actual calculations show this maximal mass is of the order of the scalar field's Compton wavelength
\begin{equation}
 M_{\rm ADM}^{\rm max}\simeq \alpha_{\rm BS} \frac{M_{\rm Pl}^2}{\mu}\simeq \alpha_{\rm BS} \, 10^{-19}M_\odot\left(\frac{\rm GeV}{\mu}\right)\ , 
\label{mini}
\end{equation}
where $M_{\odot}$ is the Sun's mass and $\alpha_{\rm BS}$ is found, numerically, to be of order of unity. Observe that the maximal mass is small, by astrophysics standards, for scalar field masses within the range of the standard model of particle physics. As such, these BSs are usually called \textit{mini-}BSs\cite{Schunck:2003kk} in the literature. 

The scalar field profile of such a BS generally has a number of nodes, or roots $n\in \mathbb{N}_0$. The $n=0$ case corresponds to fundamental states whereas $n>0$ are excited states. We shall focus on the former, since they are the most stable configurations.  Furthermore, we will study solutions with $m=1$, where $m$ is the azimuthal harmonic index. It has been shown that the maximal mass increases with increasing $m$.\cite{Liebling:2012fv,Yoshida:1997qf,Grandclement:2014msa}
This increase, however, requires  unreasonably high values of $m$, for the mass to be of order $M_{\odot}$, when $\mu$ is in the mass range of standard model particles. Ultra-light scalar fields, on the other hand, can yield astrophysical masses even for small $m$. 

A different approach to increase the maximal mass of BSs is to include self-interactions for the scalar field.
It was found by Colpi et al.\cite{Colpi:1986ye} that for spherically symmetric \textit{quartic}-BSs the maximal mass can be considerably higher:

\begin{equation}
 M_{\rm ADM}^{\rm max}\simeq 
0.062  \sqrt{\lambda}\frac{M_{\rm Pl}^3}{\mu^2}\simeq 0.062  \sqrt{\lambda}M_\odot \left(\frac{\rm GeV}{\mu}\right)^2\ .
\label{csw}
\end{equation}
Here, $\lambda>0$ is the self-interaction coupling where the self-interaction potential is given by $V(|\Psi|)=\lambda|\Psi|^4$.
It is therefore clear that by choosing a suitable $\lambda$, an astrophysically relevant mass can be obtained for BSs, even for values of $\mu$ within typical of standard model particles.

For rotating BSs, similar results were found by Ryan~\cite{Ryan:1996nk} for the large self-interaction limit, while later work considered a single value of the coupling and higher $m$ solutions.~\cite{Grandclement:2014msa,Kleihaus:2015iea}
In Ref.~\refcite{Herdeiro:2015tia} a more systematic study of rotating quartic-BSs in terms of $\lambda$ was presented.
In particular, the analogue of Eq.~\eqref{csw} was obtained for $m=1$ solutions ($cf.$ Fig.~2 in Ref.~\refcite{Herdeiro:2015tia}):
\begin{equation}
M_{\rm ADM}^{\rm max}\simeq 
0.057  \sqrt{\lambda}\frac{M_{\rm Pl}^3}{\mu^2}\simeq 0.057  \sqrt{\lambda}M_\odot \left(\frac{\rm GeV}{\mu}\right)^2\ .
\label{quartic_rotating}
\end{equation}

The purpose of this note is to expand on some physical properties of the self-interacting BSs  and Kerr BHs with scalar hair (KBHsSH) studied in Ref.~\refcite{Herdeiro:2015tia} and study whether new features appear as compared to non-self-interacting BSs and KBHsSH. A brief review of the model to be addressed is given in Sec.~\ref{sec:model}. This model contains both BSs and hairy BH solutions. In Sec.~\ref{sec:BSs}, we consider the compactness, ergo-regions, scalar field profiles and energy densities for a sample of rotating BS solutions, whose domains of existence are depicted in the top panel of Fig.~\ref{fig:figures}. The corresponding study has been performed for rotating mini-BSs in Ref.~\refcite{Herdeiro:2015gia}.
Finally, we will briefly discuss Kerr BHs with self-interacting scalar hair\cite{Herdeiro:2015tia} and their ergo-regions in Sec. \ref{sec:HBHs}.

\section{The model}
\label{sec:model}

The action of a massive complex scalar field, $\Psi$, coupled to gravity is given by
\begin{equation}
  \label{action}
	\mathcal{S} = \int d^4x \sqrt{-g}\left[\frac{R}{16\pi G}- g^{ab}\Psi^*_{,a}\Psi_{,b} - U(|\Psi|)\right] \ .
\end{equation} 
We will consider the potential
\begin{equation}
\label{pot}
U(|\Psi|)= \mu^2\left|\Psi\right|^2 + \Lambda\left|\Psi\right|^4.
\end{equation} 
where the self-coupling, $\Lambda=\lambda M_{\rm Pl}^2/\mu^2$, is positive.
The metric ansatz, for both the BSs and BHs, is given by\cite{Herdeiro:2014goa}
\begin{equation}
  ds^2 = -e^{F_0}Ndt^2 + e^{2F_1}\left( \frac{dr^2}{N} + r^2d\theta^2 \right) + e^{2F_2}r^2\sin^2\theta \left(d\varphi - Wdt \right)^2 \ ,
\end{equation}
where $N\equiv 1-r_H/r$ and $F_i,W$, $i=0,1,2$ are functions of the spherical coordinates $r$ and $\theta$ only.
Note that for BSs, $N=1$, as $r_H$, the position of the horizon, is zero.
For a scalar field ansatz, we take
\begin{equation}
\label{scalar}
  \Psi = e^{-iwt+im\varphi}\phi(r,\theta),
\end{equation}
where $w>0$ is the scalar field frequency and $m$ the azimuthal harmonic index.
%We focus on nodeless solutions with $m=1$.
%That is, the scalar field profile, $\phi(r,\pi/2)$, has no roots.

The Einstein-Klein-Gordon equations with the action Eq.~\eqref{action} result in a system of five coupled, non-linear PDEs along with two constraint equations.
The exact form of these equations and their boundary conditions can be found in Refs.~\refcite{Herdeiro:2015tia} and~\refcite{Herdeiro:2015gia}.
In those same references, one can find a description of the numerical method employed to solve these equations.

\section{Spinning boson stars with self-interaction}
\label{sec:BSs}

Boson star solutions exist along a spiral-like curve on a mass $vs.$ frequency diagram -  Fig.~\ref{fig:figures} (top panel), where five such curves for different couping values are shown. The spiral has a center but due to numerical limitations, following the spiral to its center was not possible.
The spirals shown in Fig.~\ref{fig:figures} only have two branches: a first branch from $w/\mu=1$ to a minimal frequency and a second branch from that minimal frequency to another \textit{backbending} of the curve.
A third branch is shown in Ref.~\refcite{Herdeiro:2015tia}, but we will omit them here for clarity.

\subsection{Compactness}
\label{sec:compactness}

Following Ref.~\refcite{Herdeiro:2015gia}, the inverse compactness is defined as
\begin{equation}
  \textrm{Compactness}^{-1} = \frac{R_{99}}{2M_{99}},
  \label{compactness}
\end{equation}
where $R_{99}$ is the perimeteral radius that contains $99\%$ of the BS mass, $M_{99}$.
Note that this compactness has a maximum value when the radius $R_{99}$ is equal to the Schwarzschild radius of the mass $M_{99}$, i.e. $2M_{99}$.

\begin{figure}[H]
  \begin{center}
    \includegraphics[width=0.78\textwidth]{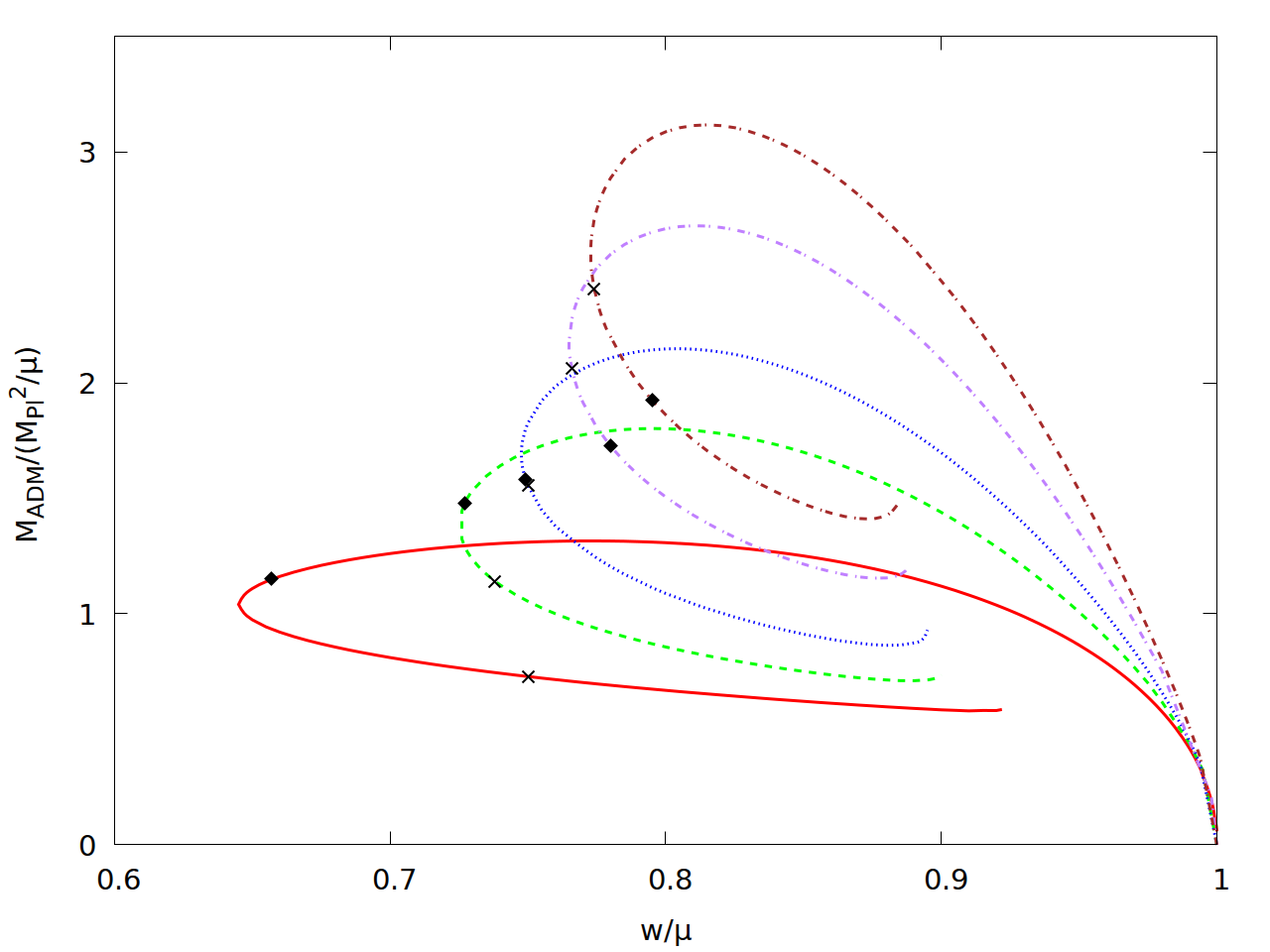}
    \includegraphics[width=0.78\textwidth]{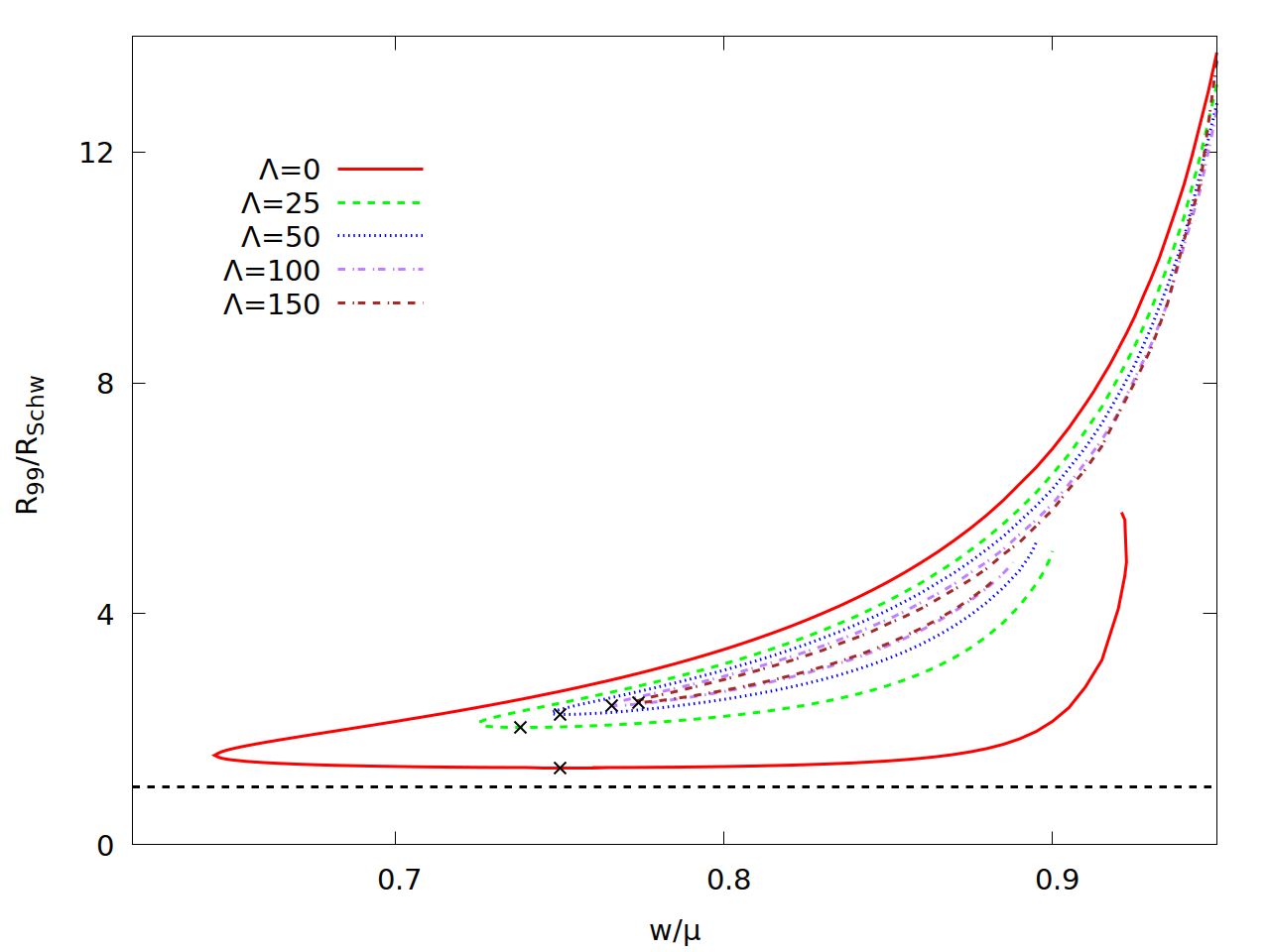}
  \end{center}
  \caption{\textbf{Top:} Mass as a function of frequency for $\Lambda=0,25,50,100,150$.
  \textbf{Bottom:} Inverse compactness as a function of frequency for the same values of $\Lambda$.
  The black dashed line shows where the compactness reaches a Schwarzschild BH.
  In both figures, the crosses show the maximum compactness solution for each coupling parameter while the black diamonds show where the first ergo-region appears, discussed in section~\ref{sec:ergo-regions}.}
  \label{fig:figures}
\end{figure}

The results for the different cases of $\Lambda$ are shown in the bottom panel of Fig.~\ref{fig:figures}.
One can see that, as one travels along the spiral (from $w/\mu=1$ and inwards) the compactness increases and this holds true for all values of $\Lambda$.
Note, however, that if one compares solutions with the same frequency but different $\Lambda$, we find that for solutions on the first branch, increasing the coupling corresponds to an increase in compactness, while for solutions on the second branch, the same increase in coupling will decrease the compactness.

There is always a solution of maximum compactness, depicted by the crosses in both the top and bottom panels of Fig.~\ref{fig:figures}.
As the self-interaction is increased, the compactness of this solution decreases. This is indeed the trend to keep in mind as a punch line: the most compact BSs are \textit{less compact} as self-interactions are added to the scalar field. This, of course, agrees with the increase in maximal mass as the self-interaction coupling is increased.

\subsection{Ergo-regions}
\label{sec:ergo-regions}

An ergo-region, defined as a spacetime region where the timelike Killing vector field (at spatial infinity) becomes spacelike, is present for a certain part of the spiral of BSs.\cite{Kleihaus:2007vk}
Ergo-regions can be understood physically as a region where physical objects, following non-spacelike worldlines, can not stay at rest when viewed by an asymptotic static observer.

We will start by studying how the appearance of an ergo-region changes, in a $(w,M)$ diagram, as the self-interaction is turned on and then increased.
In all cases, these ergo-regions are ergo-tori.\cite{Kleihaus:2007vk,Herdeiro:2014jaa}.

\begin{figure}[H]
  \begin{center}
    \includegraphics[width=0.78\textwidth]{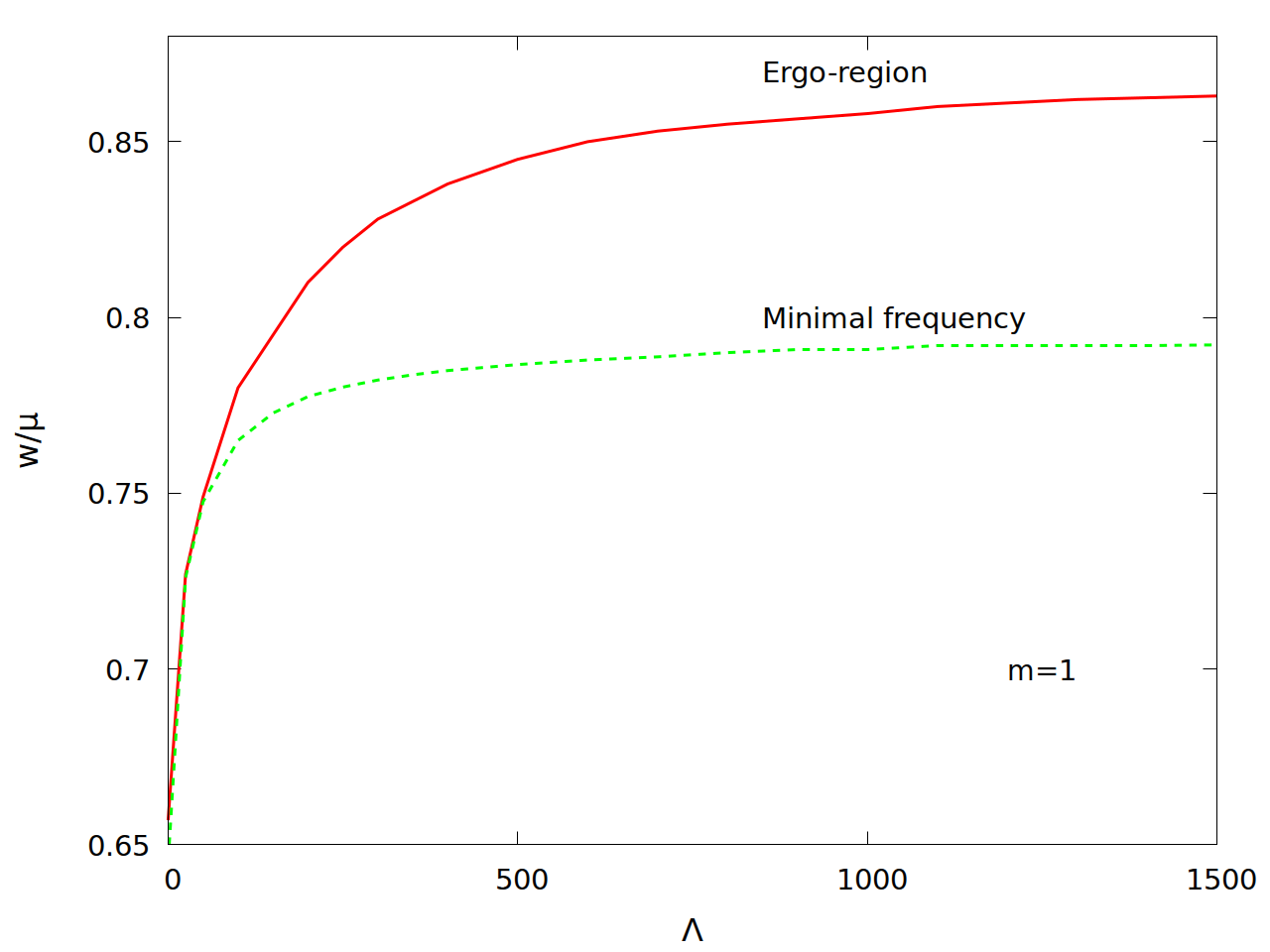}
  \end{center}
  \caption{The frequency at which the ergo-torus first appears (red solid line) and the minimal frequency (green dashed line) for self-interacting BSs for $0\leq\Lambda\leq1500$.}
  \label{fig:ergo-torus}
\end{figure}
These results are shown in the top panel of Fig.~\ref{fig:figures} as the crosses on each spiral.
For a non-self-interacting scalar field, the ergo-torus first appears for a frequency, $w_{\rm ergo}$, on the first branch after the point where the maximum mass value is reached and is present for all solutions along the spiral after that solution.
As the self-interaction is turned on, the first appearance of the ergo-torus moves along the spiral.
In Fig.~\ref{fig:ergo-torus} we show the frequency at which an ergo-torus appears (red line) as a function of the coupling parameter.
One can see that if one considers $0\leq\Lambda\leq1500$,
$w_{\rm ergo}$ seems to approach a certain limiting frequency, somewhere between $w/\mu=0.86$ and $w/\mu=0.87$ on the second branch of solutions.
In the same figure, the minimum frequency, $w_{\rm min}$, of each BS spiral (green dashed line) is shown and is seen to follow a similar pattern.

This is in agreement with the results shown in Fig.~1 of Ref.~\refcite{Herdeiro:2015tia} as the BS envelopes are shown to approach a certain frequency range in the large self-interaction limit.
This indicates that features such as the BS minimum frequency, first appearance of an ergo-region and others, converge towards some limiting value as $\Lambda$ takes very large values.

\subsection{Scalar field profiles and energy densities}
\label{sec:scalar-profiles}

\begin{figure}
  \begin{center}
    \includegraphics[width=0.78\textwidth]{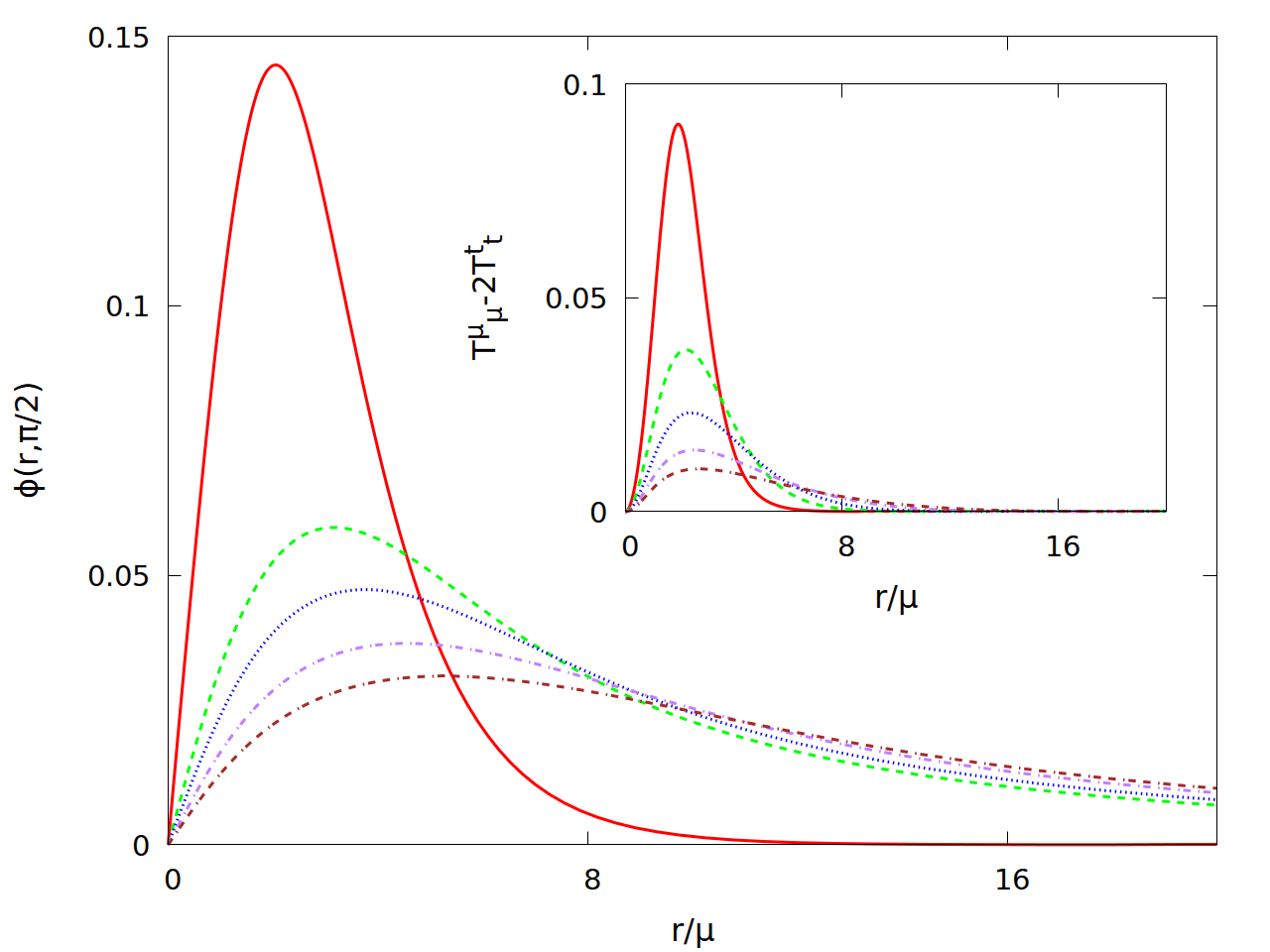}
    \includegraphics[width=0.78\textwidth]{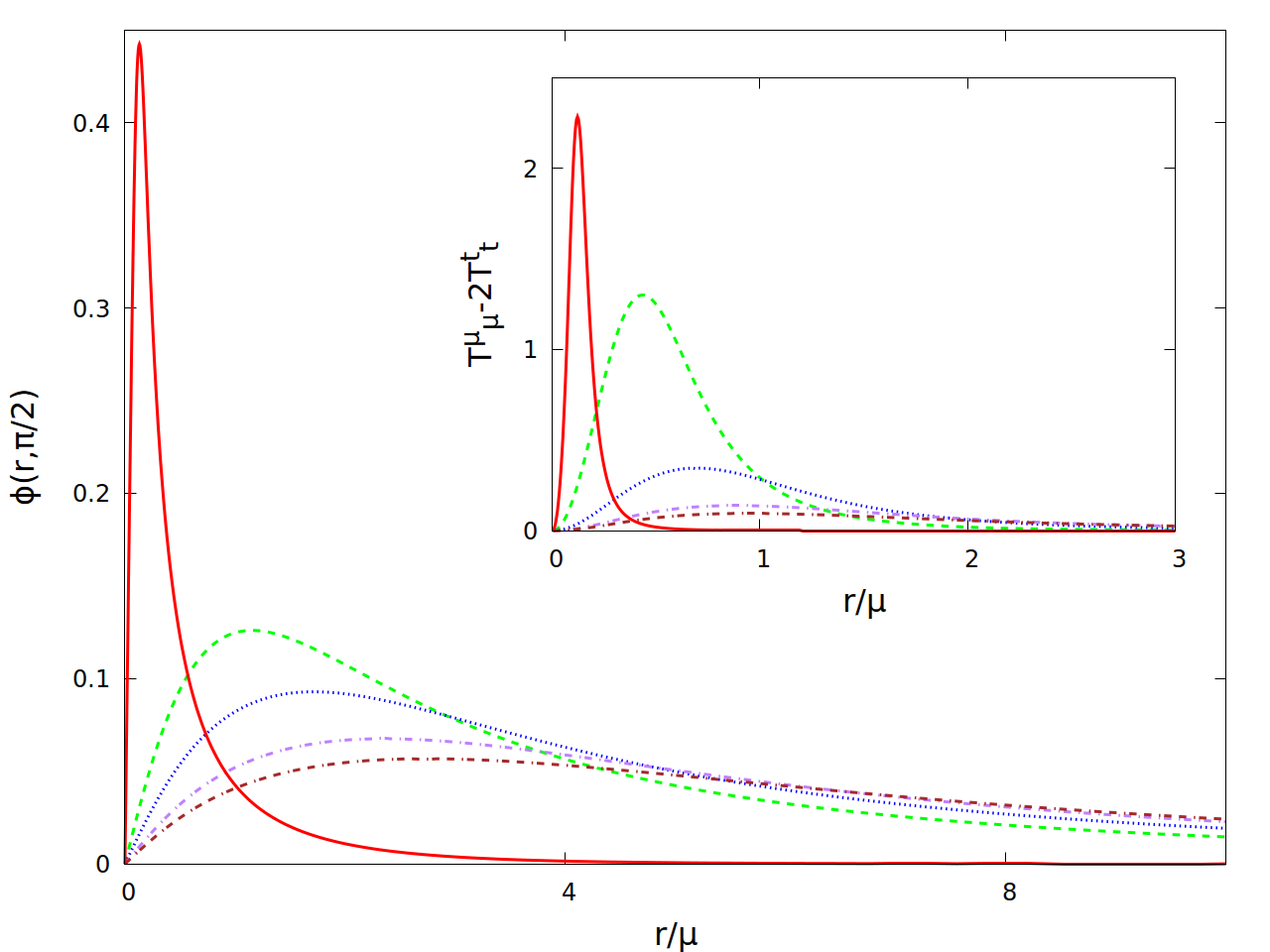}
  \end{center}
  \caption{Scalar field profiles along the equatorial plane for the maximum mass (top panel) and maximum compactness (bottom panel) solutions for various values of the self-interaction coupling: $\Lambda=0,25,50,100,150$.
  For both figures, the inset shows the energy density in the equatorial plane. Note that in the bottom panel inset, the $\Lambda=0$ curve, in solid red, is scaled by a factor of $10$.}
  \label{fig:scalar}
\end{figure}

% In Figs.~\ref{fig:scalarMaxMass} and~\ref{fig:scalarMaxCompactness} the scalar field profiles, $\phi(r,\pi/2)$,  for the maximum mass and maximum compactness  solutions for $\Lambda=0,25,50,100$ and $150$ are depicted, respectively.
In Fig.~\ref{fig:scalar} the scalar field profiles along the equatorial plane, $\phi(r,\pi/2)$, for the maximum mass (top panel) and maximum compactness (bottom panel) solutions with $\Lambda=0,25,50,100$ and $150$ are depicted.
In the inset of both figures, the energy density profile, $T^\mu_{~\mu}-2T^t_{~t}$ (also along the equatorial plane) is shown for each solution.
\footnote{In an asymptotically flat spacetime with an (asymptotically) timelike Killing vector field ($\partial_t$ in our coordinate system) this is the natural energy density defined by a Komar integral. See, e.g., the discussion in Section 6.1 of Ref.~\refcite{Herdeiro:2016tmi}.}
The maximum compactness solutions are depicted by the crosses in both the top and bottom panels of Fig.~\ref{fig:figures}.
% \begin{figure}[H]
%   \begin{center}
%     \includegraphics[width=0.75\textwidth]{BS-scalar-maxMass.png}
%   \end{center}
%   \caption{Scalar field profiles in the equatorial plane for the maximum mass solutions for various values of the self-interaction coupling: $\Lambda=0,25,50,100,150$.
%   The inset shows the energy density.}
%   \label{fig:scalarMaxMass}
% \end{figure}

As the self-interaction is turned on, the scalar field's maximal value decreases and the field decays more slowly for both cases, though the effect is larger for the maximal compactness solutions.
In the inset of both figures, the energy density of the solutions is shown to follow a similar pattern. Observe, moreover, that the difference in compactness is clearer for the bottom panel, corresponding to the maximal compactness solutions.

\section{Kerr BHs with self-interacting scalar hair}
\label{sec:HBHs}

Kerr BHs with scalar hair (KBHsSH) can be viewed as a BS with a horizon added to its centre.\cite{Herdeiro:2014goa}
It is therefore natural to ask whether one can obtain hairy BHs with self-interacting scalar hair.
Indeed, these solutions were shown to exist in Ref.~\refcite{Herdeiro:2015tia}.

\begin{figure}[H]
  \begin{center}
    \includegraphics[width=0.78\textwidth]{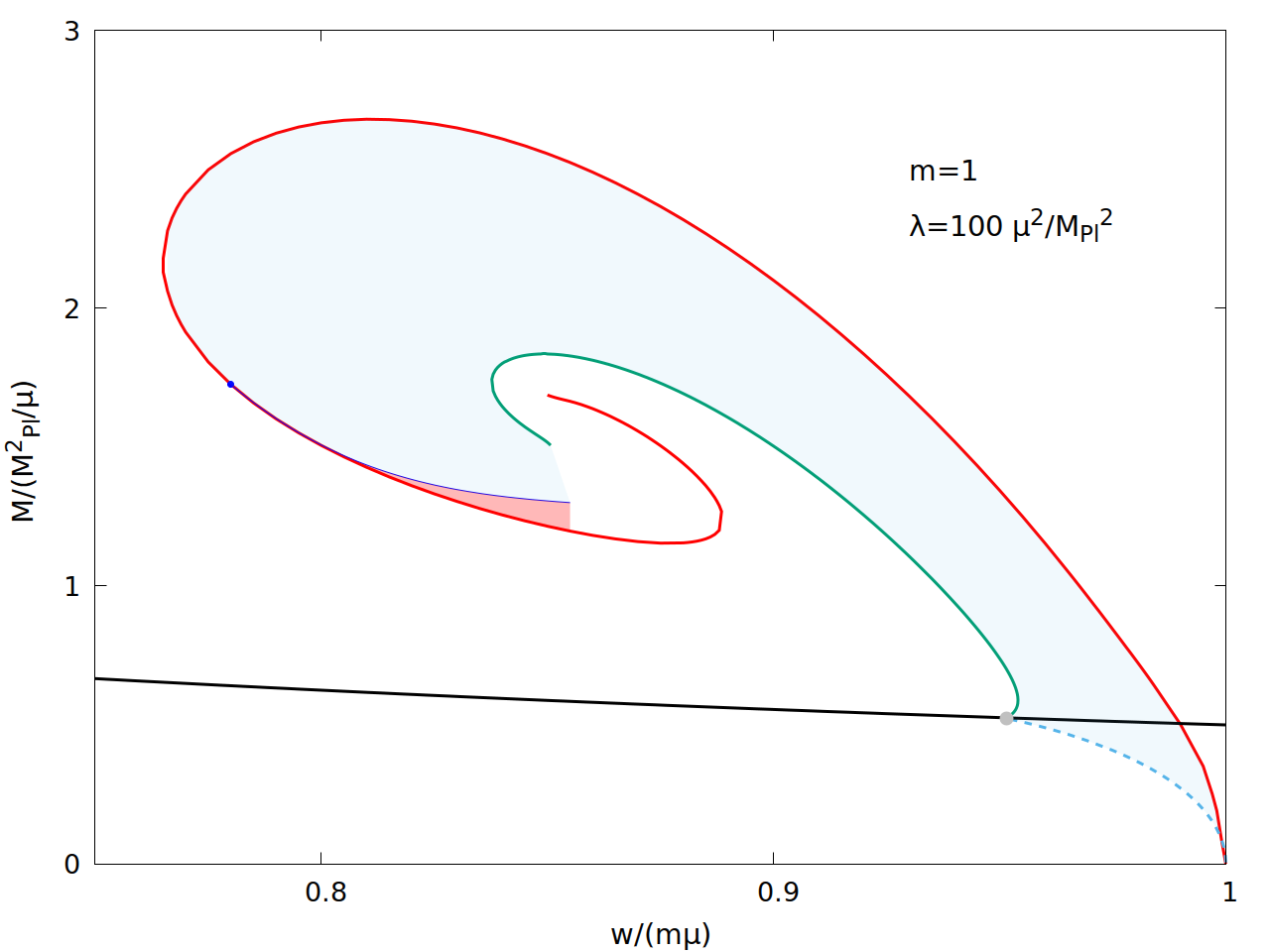}
  \end{center}
  \caption{The domain of existence of KBHsSH for $\Lambda=100$. The red curve is the boson star spiral, the green curve are the extremal hairy BHs, the blue dashed curve is the zero-mode curve, the black curve corresponds to extremal Kerr BHs and the solid blue curve and dot
  correspond to where the ergo-torus appears. The (light) blue  shaded region only has an ergo-sphere while (dark) red shaded region has an ergo-Saturn.}
  \label{fig:HBH-ergo}
\end{figure}

An interesting observation in that paper is that KBHsSH with self-interacting scalar hair, just as their BS counterparts, can become more massive than the non-self-interacting KBHsSH; yet this increase in mass arises solely from the scalar hair and not from a change in the mass of the BH at its centre.
Thus, while the ADM mass of the KBHsSH with self-interacting hair increases, their \textit{horizon mass} does not.
In fact, the horizon quantities are maximized at the so-called \textit{Hod point}\cite{Hod:2012px}, shown as a grey dot in Fig.~\ref{fig:HBH-ergo}, corresponding to the extremal Kerr BH hole obtained in the limit of vanishing hair.
Note that the Hod point does not vary with the self-interaction coupling. 
Due to this observation, these KBHsSH with self-interacting hair were said to be ``hairier but not heavier''.

As KBHsSH are continuously connected to Kerr BHs and BSs (and in fact, interpolate between them), their ergo-regions are a combination of the ergo-regions of these two objects.\cite{Herdeiro:2014jaa}

While Kerr BHs have an ergo-sphere for all values of their parameter space, BSs may possess an ergo-torus, as was seen above.
When a KBHsSH has both an ergo-sphere and an ergo-torus, it is said to have an \textit{ergo-Saturn}.\cite{Herdeiro:2014jaa}

For the non-self-interacting KBHsSH, Fig.~2 of Ref.~\refcite{Herdeiro:2014jaa} shows clearly which parts of the domain of existence of KBHsSH have an ergo-Saturn and which ones have an ergo-sphere.
As the appearance of an ergo-torus for self-interacting BSs changes with the introduction of self-interactions, it is clear that this will also affect which KBHsSH with self-interacting hair have an ergo-Saturn and which ones do not.

An extensive study of these ergo-Saturns is beyond the scope of this note. Nevertheless, we present in Fig.~\ref{fig:HBH-ergo} some results for a single value of the self-interaction coupling, $\Lambda=100$.
In the figure, the blue-covered region represents KBHsSH solutions that have an ergo-sphere but not an ergo-torus while the red-covered region shows where an ergo-torus, and thus an ergo-Saturn, appears.
The ergo-torus first appears on the 2nd branch of BS solutions for $w/\mu=0.78$ and is present thereafter.
After that point, KBHsSH solutions start possessing an ergo-torus as well but only a limited number of solutions for every fixed frequency.
Although it was not possible to fully explore the domain of existence of KBHsSH (in particular in the vicinity of the center of the BS spiral), the general pattern is suggested from Fig.~\ref{fig:HBH-ergo}.

\section{Further remarks}
\label{sec:conclusions}

The purpose of this note was to explore some new features of the self-interacting BSs and their KBHsSH counterparts by extending the results in Ref.~\refcite{Herdeiro:2015tia}.
It was found that these BSs are, in general, less compact than the non-self-interacting (mini-)BSs.
This conclusion was reached in three different ways:
i) First, the compactness as defined in Sec.~\ref{sec:compactness} is less than that of their non-self-interacting counterparts (for the most compact BSs).
ii) Next, the appearance of an ergo-torus happens further along the frequency-mass spiral as the self-interaction coupling increases (though there is some saturation as one reaches very high coupling).
iii) Lastly, the scalar field profiles in the equatorial plane decrease in maximal amplitude and spread out as the self-coupling increases.
This been shown for both the maximal mass and maximal compactness solutions.
The same behavior was observed for the energy densities of those same solutions.

Intuitively, this decrease in compactness trend comes as no surprise, as the positive quartic-self-interaction term acts as a repulsive force for the field.
However, the slight \textit{increase} in compactness along a part of the first branch of solutions for a given coupling parameter indicates that the picture is not be as simple as one first assumes.

As expected, these qualitative changes are inherited by the KBHsSH with self-interactions.
These BHs have been shown to be more massive than their mini-KBHsSH counterparts.
This increase in mass however, is only due to the scalar field and its self-interactions, while the maximal horizon mass does not change.\cite{Herdeiro:2015tia}
As with the ergo-regions of the self-interacting BSs we have found that the the parts of the domain of existence that exhibit ergo-Saturns moves closer to the centre of the spiral.

The results presented here suggest that both BSs and KBHsSH with self-interactions can have interesting, and distinct, phenomenology as compared to their non-self-interacting counterparts.
As was shown in Ref.~\refcite{Cunha:2015yba}~, the gravitational lensing of BSs is intricate and when those BSs are combined with BHs, shadows that are distinct from the classical Kerr BH shadows are found.
As such, we anticipate different gravitational lensing patterns for the self-interacting BSs and different shadows for KBHsSH with self-interacting scalar hair. Study of these features is in progress.

Self-interactions can also affect the interaction between scalar fields and BHs in different ways.
One example shows up at the linear level when one considers which Kerr BHs can support a stationary scalar field without backreaction, leading to \textit{stationary scalar clouds}.\cite{Hod:2012px}
For a free complex scalar field, this only occurs for a one-dimensional line of Kerr BHs, for a given scalar field mode  (seen in Fig.~\ref{fig:HBH-ergo} as the blue dashed curve).
By contrast, if one adds self-interactions of the type
$$V(|\Psi|)= \mu^2\left|\Psi\right|^2 - \beta\left|\Psi\right|^4 + \alpha\left|\Psi\right|^6,$$ 
where $\alpha,\beta>0$, to these clouds, the one-dimensional line of Kerr BHs that can support clouds is expanded into a two-dimensional plane.
Thus, a larger set of Kerr BHs can support these, so-called, \textit{Q-clouds}.\cite{Herdeiro:2014pka}
A preliminary study of these Q-clouds in the fully non-linear regime is presented in Ref.~\refcite{Herdeiro:2015tia}.

Finally, it would be interesting to study the quadrupole moments, orbital frequency at the ISCO and other phenomenological quantities for these BHs with self-interacting scalar field hair and see how they compare to both the ``classical'' Kerr BHs and KBHsSH.
Since BSs can have quadrupole moments that are much higher than Kerr BHs, the self-interacting solutions, with their increased mass, are expected to have even higher quadrupole moments.
This expected increase due to self-interactions would most likely be inherited by the quadrupole moments of self-interacting KBHsSH.
Due to the reduced compactness of these solutions, one would expect that KBHsSH with self-interactions follow a similar pattern.
Thus, we also expect that the orbital frequency at the ISCO to be lower, as the ISCO will most likely move away from the centre of the BH.

\section{Acknowledgements}

C. H. and E. R. acknowledge funding from the FCT-IF programme.
H.R. is supported by the grant PD/BD/109532/2015 under the MAP-Fis Ph.D. programme.
This work was partially supported by the H2020-MSCA-RISE-2015 Grant No. StronGrHEP-690904, and by the CIDMA project UID/MAT/04106/2013.
Computations were performed at the Blafis cluster, in Aveiro University.

%\begin{thebibliography}{000} %for 3 digits
%\begin{thebibliography}{00}  %for 2 digits

\end{document}